\documentclass[aps,prl,twocolumn,showpacs,superscriptaddress]{revtex4-1}

\usepackage{graphicx}
\usepackage{color}
\usepackage{txfonts}

\newcommand{\NIO}{Nd$_2$Ir$_2$O$_7$}
\newcommand{\tn}{T$_N$}
\begin{document}

\title{Direct determination of the spin structure of Nd$_2$Ir$_2$O$_7$ by means of neutron diffraction}

\author{H. Guo}
\affiliation{Max-Planck-Institute for Chemical Physics of Solids, N\"{o}thnitzer Str. 40, D-01187 Dresden, Germany}

\author{C. Ritter}
\affiliation{Institut Laue-Langevin, Boite Postale 156X, F-38042 Grenoble Cedex 9, France}

\author{A. C. Komarek}
\email[]{Komarek@cpfs.mpg.de}
\affiliation{Max-Planck-Institute for Chemical Physics of Solids, N\"{o}thnitzer Str. 40, D-01187 Dresden, Germany}

\date{\today}

\begin{abstract}
We report on the spin structure of the pyrochlore iridate \NIO\ that could be directly determined by means of powder neutron diffraction. Our magnetic structure refinement unravels a so-called all-in/all-out magnetic structure that appears in both, the Nd and the Ir sublattice. The ordered magnetic moments  at 1.8~K amount to 0.34(1)~$\mu_\mathrm{B}$/Ir$^{4+}$ and 1.27(1)~$\mu_\mathrm{B}$/Nd$^{3+}$. The Nd$^{3+}$ moment size at 1.8~K is smaller than that expected for the Nd$^{3+}$ ground state doublet. 
On the other hand, the size of the ordered moments of the  Ir$^{4+}$ ions at 1.8~K agrees very well with the value expected for a $J_\mathrm{eff}$~=~1/2 state based on the presence of strong spin-orbit coupling in this system. Finally, our measurements reveal a parallel alignment of the Nd$^{3+}$ moments with the net moment of its six nearest neighboring Ir$^{4+}$ ions.
\end{abstract}

\pacs{75.25.-j, 75.30.Gw, 75.47.Lx, 71.70.Ej}
\maketitle

The 5$d$ transition metal oxides have attracted substantial attention due to the interplay between the relatively large spin-orbit coupling (SOC) and electron-electron correlations ($U$), which may result in exotic electronic phases such as topological Mott insulators, Weyl semimetals and axion insulators \cite{Wan_theory,Pesin-theory,Yang-theory, William-theory,Shinaoka_theory,weylA,weylB}.
The pyrochlore iridates ($R_2$Ir$_2$O$_7$, $R$ = rare earth and Y) are a fertile playground to realize these topological phases. For the pyrochlore structure, the ions at the $R$ site or Ir site form a network of corning-sharing tetrahedra with the two sublattices penetrating each other. These pyrochlore iridates exhibit a metal-insulator transition (MIT) at $T_\mathrm{MI}$ which continuously decreases with increasing $R$-ionic radius. The MIT disappears ($T_\mathrm{MI}$~=~0)  between $R$ = Nd and Pr \cite{Matsuhira-2007,Matsuhira_transport,Nakatsuji-Pr-227}.
No thermal hysteresis effect can be observed at $T_\mathrm{MI}$ (\tn), thus, indicating a second order transition \cite{Matsuhira_transport}. At \tn, magnetization, $\mu$SR and resonant X-ray scattering experiments suggest an ordering of the Ir$^{4+}$ moments \cite{Matsuhira_transport,Guo_musr,Zhao_Eu,Disseler_Y,Sagayama_xray}. At lower temperatures, the $d-f$ interaction may induce the magnetic ordering at the $R$ site (also depending on the single-ion anisotropy at the $R$ site \cite{Neutron_anisotropy}).

Among these pyrochlore iridates, \NIO\ (with $T_\mathrm{MI}\sim$30~K) is of particular interest. Recent experimental studies combined with theoretical calculations show that this compound exhibits highly anisotropic magnetoresistance despite the unchanged cubic symmetry \cite{structure} above and below $T_\mathrm{MI}$ \cite{Tian_nat,Ueda_prl}.
It was shown that a quantum MIT is realized when a magnetic field of $\sim$10~T is applied only along the [0~0~1] direction, with the ground state changing from an insulating magnetic ordered state to a semimetal state \cite{Tian_nat,Ueda_prl}.
This could be explained by a reconstruction of the band structure concomitant with the change of the magnetic structure in the Nd sublattice from an all-in/all-out (AIAO) to a two-in/two-out magnetic structure.

Several theoretical calculations \cite{Wan_theory,Pesin-theory,Yang-theory,William-theory,Shinaoka_theory,weylB} have demonstrated that the AIAO magnetic structure has the lowest energy when $U$ becomes larger than a critical value $U_c$. The AIAO magnetic ordering is also indispensable for realizing the topologically nontrivial Weyl semimetal state \cite{Wan_theory,Pesin-theory,Yang-theory, William-theory,Shinaoka_theory,weylB}.
Therefore, it is essential to unambiguously determine the magnetic structure of \NIO\ experimentally in order to understand these emergent phenomena in pyrochlore iridates.
Apart from the Nd sublattice where an AIAO magnetic structure was observed, the direct evidence for an AIAO structure at the Ir sublattice from neutron scattering is lacking for all pyrochlore iridates \cite{Tomiyasu_neutron}.
Despite enormous effort, the magnetic structure of the Ir sublattice could not be measured directly in neutron scattering experiments due to the elevated neutron absorption of the Ir ion and the small ordered magnetic moments of the Ir ions.
To the best of our knowledge, there exists only evidence for an AIAO structure of the Ir sublattice suggested by resonant X-ray scattering techniques  in Eu$_2$Ir$_2$O$_7$ and in Sm$_2$Ir$_2$O$_7$  \cite{Sagayama_xray,Sm}.

Besides the Ir sublattice in \NIO\ even the moment sizes at the Nd sublattice are still under debate. From PND, Tomiyasu \textit{et al.} suggested an AIAO structure of the Nd sublattice with moment sizes that amount to $\sim$1.3~$\mu_\mathrm{B}$/Nd at 9~K and $\sim$2.3~$\mu_\mathrm{B}$/Nd at 0.7~K \cite{Tomiyasu_neutron}. However, the statistics seems insufficient for an unambiguous determination of the Nd moments in Ref. \cite{Tomiyasu_neutron}.
Subsequent studies of \NIO\ by means of $\mu$SR measurements yield controversial results. One $\mu$SR study supports the AIAO magnetic structure on both sublattices \cite{Guo_musr} whereas the other one indicates that the Nd sublattice can not order in an AIAO magnetic structure since the calculated internal field at the muon site is much larger than the experimental value with the proposed magnetic structure \cite{Disseler_muon_site}.

Here, we report the first successful determination of the magnetic structure of a pyrochlore iridate by means of neutron diffraction.
Using an optimized sample geometry and a strong neutron flux we were able to solve the magnetic structure of \NIO\ including the spin structure of the Ir sublattice. Our results unambiguously show that both the Ir and the Nd sublattices order in the AIAO structure, with the magnetic moments of 0.21(1)~$\mu_\mathrm{B}$/Ir and 0.27(1)~$\mu_\mathrm{B}$/Nd at 15~K and 0.34(1)~$\mu_\mathrm{B}$/Ir and 1.27(1)~$\mu_\mathrm{B}$/Nd at 1.8~K. Moreover, we find that the Nd moment and the net moment of all six nearest neighboring Ir ions are aligned parallel (ferromagnetic).

Polycrystalline samples of \NIO\ were synthesized by solid state reaction as described in Ref.~\cite{Matsuhira_transport}. The phase purity was confirmed by X-ray diffraction (XRD) and PND. PND measurements were performed at the D20 and the D2B diffractometer at the Institut Laue-Langevin (ILL) in Grenoble, France.
For the magnetic structure determination and for nuclear structure refinement we have chosen an incident neutron wave length of 2.41~$\AA$ (D20 diffractometer)  and 1.594~$\AA$ (D2B diffractometer). In order to reduce the impact of neutron absorbtion effects by the Ir atoms we have filled the powder sample (of about 7~g mass) into a hollow vanadium cylinder.

\begin{figure}
\centering
\includegraphics[width=1\columnwidth]{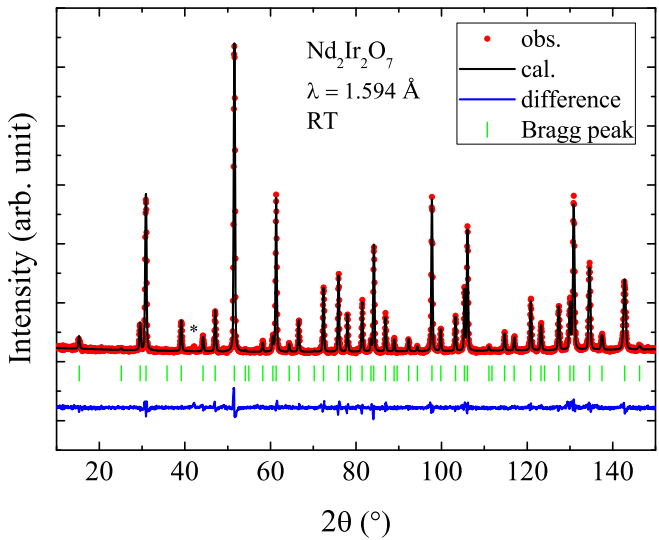}
\caption{(color online) PND pattern of \NIO\ measured at room temperature at the D2B diffractometer together with a Rietveld fit ($a$~$=$~10.3683(2)~$\AA$). \emph{Red dots:} experimentally observed intensities, \emph{black curve:} calculated intensities, \emph{blue curve:} difference between experimentally observed and calculated intensities. \emph{Green bars:} Bragg peak positions. The asterisk indicates a small impurity peak from IrO$_2$. The refinable parameter related to the oxygen position at site 48$f$ ($x$, 0.125, 0.125) amounts to $x$ = 0.3315(2). The final reliability factors are $R_B$ = 3.87 \% and $\chi^2$ = 2.35.}
\label{RT_pattern}
\end{figure}

The Rietveld refinement of the nuclear structure of \NIO\ is shown in Figure~\ref{RT_pattern}. A refinement with space group $Fd\bar{3}m$ describes the measured data well. Besides that a tiny impurity phase (less than 1~\%) of IrO$_2$ is observable in the data. Our refinement shows that no stuffing (defect fluorite structure) appears in our material. Moreover, a refinement of the oxygen occupancy indicates no oxygen off-stoichiometry within the accuracy of our measurements,
i.e. indicate an oxygen content of about 6.99(3).

\begin{figure}
\centering
\includegraphics[width=1\columnwidth]{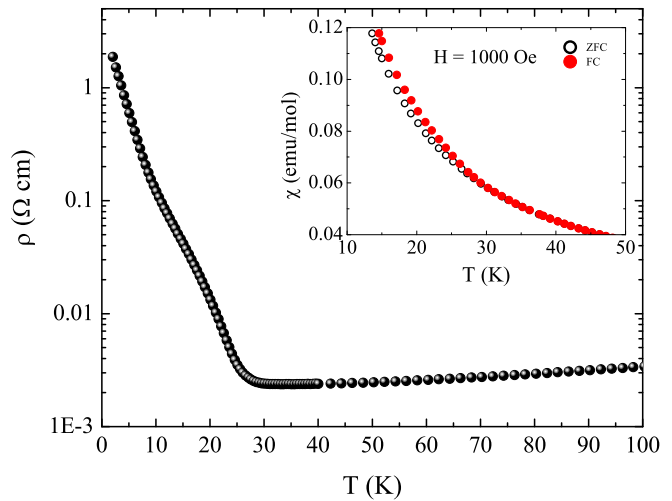}
\caption{Temperature dependence of resistivity. The metal-insulator transition temperature is about 30 K. The inset shows the magnetic susceptibility measurement at $H$ = 1000 Oe. A small bifurcation between the field-cooled and zero-field-cooled is also observed at about 30 K.}
\label{resistivity}
\end{figure}

Our sample was also characterized by resistivity and magnetic susceptibility measurements. As shown in Fig.~\ref{resistivity} a MIT takes place at $\sim$~30~K, below which the resistivity increases by roughly three orders of magnitude on cooling from 30~K to 2~K, thus, indicating a good sample quality compared to others reported in literature \cite{Matsuhira_transport,Guo_musr}. In the vicinity of the MIT, the magnetic susceptibility exhibits a bifurcation between the field-cooled (FC) and zero-field-cooled (ZFC) measurements. This is also consistent with previous results.

\begin{figure*}
\centering
\includegraphics[width=2\columnwidth]{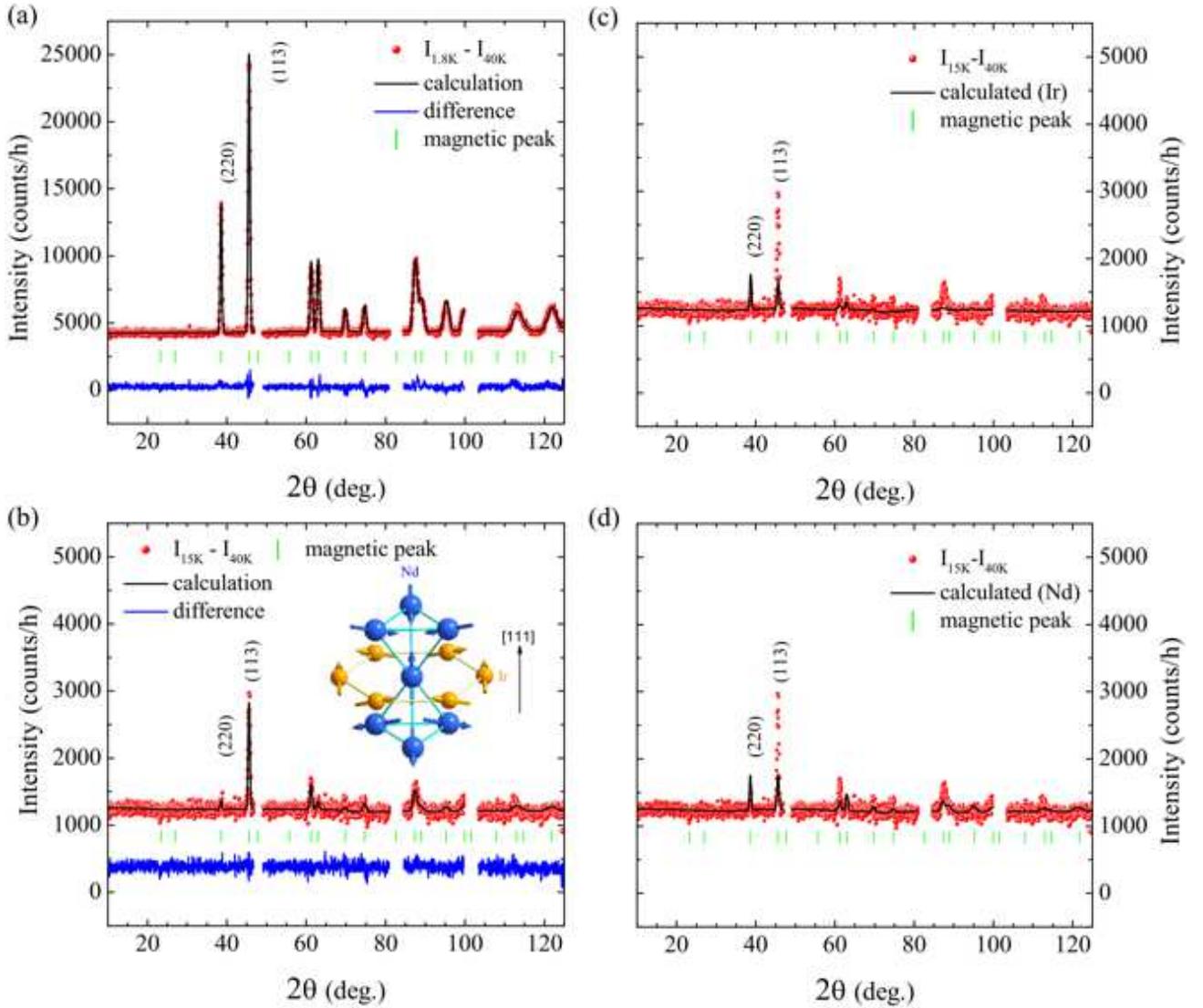}
\caption{Rietveld refinement of the magnetic structure according to IR $\Gamma_3^1$ for the difference pattern (a) between 1.8~K and 40~K and (b) between 15~K and 40~K. This data was measured at the D20 diffractometer using a neutron wave length of $\lambda$~=~2.41~$\AA$. The red dots are experimental data, black and blue curves are the calculated pattern and the difference between experimental and calculated values. Green bars indicate the magnetic peak positions. The regions around strong Bragg peaks were excluded from the refinement due to the shift of lattice constants. The inset of (b) shows the AIAO magnetic structure of the Nd sublattice with six surrounding Ir ions. The moment size of Ir ions compared to that of Nd ions are enlarged for clearance.
(c,d) Magnetic pattern at 15~K refined by assuming the ordering only at (c) the Ir-sublattice or only at (d) the Nd-sublattice (according to IR $\Gamma_3^1$).}
\label{magnetic_pattern}
\end{figure*}

Our $\mu$SR experiment on \NIO\ indicate that the Ir$^{4+}$ moments order below $\sim$30~K and become saturated below about 20~K. The Nd$^{3+}$ moments order below about 9~K \cite{Guo_musr}. Therefore, we measured \NIO\ at 40~K, 15~K and 1.8~K at the D20 diffractometer with high neutron flux and long counting times.
In Fig.~\ref{magnetic_pattern}(a) the difference between the 1.8~K and 40~K data is shown. All the magnetic peaks can be indexed by a propagation vector $\textbf{k}$~=~0. Magnetic symmetry analysis has been performed using the \textit{FULLPROF} program package \cite{Fullprof}.
For the space group $Fd\bar{3}m$ and for $\textbf{k}$~=~(0,~0,~0) the magnetic reducible representation $\Gamma_{mag}$ for the Nd$^{3+}$ ($16d$) site and the Ir$^{4+}$ ($16c$) site is decomposed as a direct sum of four nonzero irreducible representations (IRs):
\begin{equation}\label{}
  \Gamma_{mag} = \Gamma_3^1 \oplus \Gamma_6^2 \oplus \Gamma_8^3 \oplus 2\Gamma_{10}^3
\end{equation}

The basis vectors for these IRs are listed in Tab.~\ref{IRs}. For the magnetic structure refinement, the nuclear structure was refined at 40~K in order to obtain the scale factor which was, then, fixed for the subsequent magnetic structure refinements. As shown in Fig.~\ref{magnetic_pattern}(a,b) the measured magnetic intensities can be fitted only by using the IR $\Gamma_3^1$ which corresponds to the so-called AIAO magnetic structure in the pyrochlore lattice. All other IRs are not able to describe our neutron data, see magnetic R-values in Tab.~\ref{IRs}. Since there is only one basis vector for the one-dimensional IR $\Gamma_3^1$ only one coefficient $c_i$ ($i$ = Nd and Ir) has to be refined for each magnetic sublattice, and, the refinement of the 1.8~K data yields an ordered magnetic moment of 0.34(1)~$\mu_\mathrm{B}$ at the Ir and 1.27(1)~$\mu_\mathrm{B}$ at the Nd site. The reliability factor amounts to $R_{mag}$ = 5.76\%. At 15~K the ordered moments amount to 0.21(1)~$\mu_\mathrm{B}$/Ir and 0.27(1)~$\mu_\mathrm{B}$/Nd.
The obtained magnetic structure is shown in the inset of Fig.~\ref{magnetic_pattern}(b).
The Nd moment and the net moment of its six nearest neighboring Ir ions are aligned parallel (ferromagnetic) .
Our results are consistent with our findings in $\mu$SR data \cite{Guo_musr} which indicate that the Ir$^{4+}$ moments are almost saturated below 20~K while the strongest increase of the ordered moment of the Nd$^{3+}$ ions appears below about 9~K.

We would like to stress that a refinement of our neutron data with an AIAO magnetic structure where only one sublattice orders is not successful, see Fig.~\ref{magnetic_pattern}(c,d).
\begin{table*}
\caption{Nonzero IRs together with basis vectors $\psi_\nu$ for Nd and Ir atoms in \NIO\ with space group $Fd\bar{3}m$ and \textbf{k}~=~\textbf{0} propagation vector obtained from representational analysis. The atoms of the nonprimitive basis are defined as Nd\#1: (0.50, 0.50, 0.50), Nd\#2: (0.25, $-$0.25, 1.00), Nd\#3: ($-$0.25, 1.00, 0.25), Nd\#4: (1.00, 0.25, $-$0.25), Ir\#1: (0, 0, 0), Ir\#2: (0.75, 0.25, 0.50), Ir\#3: (0.25, 0.50, 0.75), Ir\#4: (0.50, 0.75, 0.25).
Additionally, the magnetic R-values (R$_{mag}$) from our best fit for each IR is listed. Only the R$_{mag}$ value for the refinement based on IR $\Gamma_3^1$
for both sublattices yields satisfactory low values, thus, indicating the appearance of an AIAO magnetic structure for the Ir sublattice.  \label{IRs}}
\begin{ruledtabular}
\begin{tabular}{lccccccc}
IRs & R$_{mag}$ & $\psi_\nu$ & component & atom\#1 &  atom\#2 &  atom\#3 &  atom\#4 \\
 \hline
$\Gamma_3^1$ & 5.76\% & $\psi_1$ & Real &(1\,1\,1) & ($-$1\,$-$1\,1) & ($-$1\,1\,$-$1) & (1\,$-$1\,$-$1) \\

$\Gamma_6^2$ & 55.4\% & $\psi_1$ & Real      & (1\,$-$0.5\,$-$0.5)& ($-$1\,0.5\,$-$0.5) &  ($-$1\,$-$0.5\,0.5) & (1\,0.5\,0.5) \\
             & &          & Imaginary & (0\,$-$0.87\,0.87) & (0\,0.87\,0.87) & (0\,$-$0.87\,$-$0.87) & (0\,0.87\,$-$0.87) \\
             & & $\psi_2$ & Real      &(0.5\,$-$1\,0.5) & ($-$0.5\,1\,0.5) & ($-$0.5\,$-$1\,$-$0.5) & (0.5\,1\,$-$0.5) \\
             & &          & Imaginary &(0.87\,0\,$-$0.87) & ($-$0.87\,0\,$-$0.87) & ($-$0.87\,0\,0.87) & (0.87\,0\,0.87) \\
$\Gamma_8^3$ & 82.6\% & $\psi_1$ & Real      & (1\,$-$1\,0)& ($-$1\,1\,0)& (1\,1\,0) & ($-$1\,$-$1\,0)\\
             & & $\psi_2$ & Real      & (0\,1\,$-$1)& (0\,1\,1) & (0\,$-$1\,$-$1) & (0\,$-$1\,1)\\
             & & $\psi_3$ & Real      & ($-$1\,0\,1) & ($-$1\,0\,$-$1) & (1\,0\,-1)& (1\,0\,1)  \\	
$\Gamma_{10}^3$ & 55.0\% & $\psi_1$ & Real     & (1\,1\,0) & ($-$1\,$-$1\,0) & (1\,$-$1\,0)& ($-$1\,1\,0)\\
             & & $\psi_2$ & Real      & (0\,0\,1) & (0\,0\,1) & (0\,0\,1) & (0\,0\,1)\\
             & & $\psi_3$ & Real      & (0\,1\,1) & (0\,1\,$-$1) & (0\,$-$1\,1) & (0\,$-$1\,$-$1)\\
             & & $\psi_4$ & Real      & (1\,0\,0)&  (1\,0\,0)& (1\,0\,0) & (1\,0\,0)\\
             & & $\psi_5$ & Real      & (1\,0\,1)& (1\,0\,$-$1)& ($-$1\,0\,$-$1)& ($-$1\,0\,1)\\
             & & $\psi_6$ & Real      & (0\,1\,0) & (0\,1\,0) & (0\,1\,0) & (0\,1\,0) \\
\end{tabular}
\end{ruledtabular}
\end{table*}
Especially the intensity of the (2~2~0) and (1~1~3) peaks would be similar in such models which is in strong contrast to our experimental observations. At 15~K where Nd and Ir moment sizes are of more similar size (i.e.  0.21(1)~$\mu_\mathrm{B}$/Ir and 0.27(1)~$\mu_\mathrm{B}$/Nd) and, hence, the interference effects larger, this discrepancies are most apparent, see Fig.~\ref{magnetic_pattern}(c,d). But, even at 1.8~K where the Nd moment size is dominating our measurement data can not be described by a model that contains only spins at one sublattice.
This unambiguously corroborates our observation of an additional ordered magnetic moment at the Ir site.
In previous PND measurements of Y$_2$Ir$_2$O$_7$ with nonmagnetic Y$^{3+}$ ions no indication for the ordering from Ir$^{4+}$ moments has been found \cite{Shapiro_Y} which shows the difficulty in detecting these small ordered Ir moments and the importance of our findings in \NIO.

For $D_{3h}$ symmetry the crystal electric field (CEF) effect splits the Nd$^{3+}$ $J$ = 9/2 multiplet into five Kramers doublets. The ground state doublet is highly anisotropic with magnetic moments pointing to the center of the tetrahedron. From CEF analysis the estimated ground state moment amounts to $\sim$2.37~$\mu_\mathrm{B}$/Nd \cite{neutron-crystal-field}. With 1.27(1)~$\mu_\mathrm{B}$/Nd our observed ordered Nd moment is only about half of that size expected for the CEF ground state.
In contrast to our observations previous studies of the Nd sublattice \cite{Tomiyasu_neutron} report ordered Nd moments of the size of 2.3$\pm$0.4~$\mu_\mathrm{B}$/Nd at 0.7~K. However, the statistics in Ref.~\cite{Tomiyasu_neutron} seems not to be sufficient for a reliable determination of the accurate moment sizes.
The very good statistics of our data - see Fig.~\ref{magnetic_pattern} - enables us to refine the magnetic structure of \NIO\ accurately.
Note, that $\mu$SR experiments indicate a fluctuating Nd moment \cite{Guo_musr}.
Moreover, a reduction of the Nd$^{3+}$ ordered moments has been also observed in Nd$_2$Hf$_2$O$_7$ (0.62(1)~$\mu_\mathrm{B}$/Nd at 0.1~K) \cite{Anand_BV} and in Nd$_2$Sn$_2$O$_7$ (1.708(3)~$\mu_\mathrm{B}$/Nd at 0.06~K) \cite{Bertin_Nd2Sn2O7}. These two latter compounds also exhibit an AIAO magnetic order for the Nd sublattice, and, the reduction of the Nd$^{3+}$ moments was attributed to strong quantum fluctuations \cite{Anand_BV}.
Furthermore, the magnetic structure of the Nd sublattice as proposed in Ref.~\cite{Tomiyasu_neutron} was also questioned in other $\mu$SR studies \cite{Disseler_muon_site}.
Our results unambiguously show that the Nd sublattice orders in the AIAO structure but with reduced Nd$^{3+}$ moment sizes.
More detailed calculations of the muon site and the interaction between muon and surrounding moments will be needed in order to explain the internal field observed by $\mu$SR experiments.
The ordered magnetic moment of the Ir$^{4+}$ ions which amounts to 0.34(1)~$\mu_\mathrm{B}$/Ir at 1.8~K is consistent with a SOC based $J_\mathrm{eff}$ = 1/2 state picture with a predicted moment of $\sim$1/3~$\mu_\mathrm{B}$/Ir~\cite{Kim-Sr2IrO4,Shapiro_Y}.

Concluding, we have succeeded in unambiguously determining the spin structure of \NIO\ by means of PND and were able to show that both the Nd and Ir sublattices order in the AIAO magnetic structure. The Nd$^{3+}$ ordered moments amount to 1.27(1)~$\mu_\mathrm{B}$ at 1.8~K which is smaller than the expected value for the CEF ground state doublet.
Moreover, the size of Ir$^{4+}$ ordered moments of 0.34(1)~$\mu_\mathrm{B}$ at 1.8~K  agrees very well with a SOC based $J_\mathrm{eff}$ = 1/2 state.

\par We acknowledge L.~H.~Tjeng, P.~Thalmeier, O.~Stockert and C.~Geibel for helpful discussions.

\begin{acknowledgments}

\end{acknowledgments}

\bibliography{reference}

\end{document}